\documentclass[twocolumn, amsmath, amssymb, aps, prb]{revtex4-1}

\usepackage{graphicx}
\usepackage[space]{grffile}
\usepackage{dcolumn}
\usepackage{bm}
\usepackage{subfigure}
\usepackage{mathrsfs}
\usepackage{booktabs}
\usepackage{notoccite}
\usepackage{float}
\usepackage{placeins}
\usepackage{enumerate}
\usepackage{gensymb}
\usepackage{xcolor}

\begin{document}

\title{Voltage-dependent cluster expansion for electrified solid-liquid interfaces:\\ Application to the electrochemical deposition of transition metals}

\author{Stephen E. Weitzner$^*$}
\author{Ismaila Dabo}
\affiliation{Department of Materials Science and Engineering, Materials Research Institute, and Penn State Institutes of Energy and the Environment, The Pennsylvania State University, University Park, PA 16802, USA \\  Email: weitzner@psu.edu}

\begin{abstract}
The detailed atomistic modeling of electrochemically deposited metal monolayers is challenging due to the complex structure of the metal-solution interface and the critical effects of surface electrification during electrode polarization. Accurate models of interfacial electrochemical equilibria are further challenged by the need to include entropic effects to obtain accurate surface chemical potentials. We present an embedded quantum-continuum model of the interfacial environment that addresses each of these challenges and study the underpotential deposition of silver on the gold (100) surface. We leverage these results to parameterize a cluster expansion of the electrified interface and show through grand canonical Monte Carlo calculations the crucial need to account for variations in the interfacial dipole when modeling electrodeposited metals under finite-temperature electrochemical conditions.
\end{abstract}

\maketitle

\section{Introduction}
The underpotential deposition (UPD) of transition metal ions is an effective and widely applicable method to determine the active surface area of electrodes, to perform controlled galvanic replacement reactions for the deposition of noble metals, as well as to control the shape and architecture of metallic nanoparticles for catalysis, sensing, and biomedical applications.\cite{Aoun2003,Aldana-Gonzalez2015,Maksimov2017,Price2011,Yan2015,Personick2011,Yu2009,Personick2011} In this interfacial process, metal cations are reduced and adsorbed to the surface of a more noble metal forming a stable partial- to full-monolayer at voltages more positive than the reduction potential of the cation.\cite{Kolb1974} 

First principles density functional theory (DFT) has been applied to obtain atomistic insights into the stability and structure of the metal monolayers achieving varying degrees of correspondence with experimental voltammetry.\cite{Sanchez1999,Sanchez2001,Sanchez2002,Karlberg2007,Greeley2010,Gimenez2010,Velez2012} These calculations are typically performed in the absence of a solvent; however, key features of the interface such as anion co-adsorption have been included when warranted, leading to enhanced descriptions of the interface.\cite{Gimenez2010,Velez2012} Entropic effects have additionally been considered to obtain surface chemical potentials by including ideal configurational entropy or by fitting an Ising-like Hamiltonian to DFT results and subsequently performing grand canonical Monte Carlo calculations. These approaches have been applied to study the UPD of hydrogen on platinum surfaces at finite temperatures, underscoring the importance of configurational entropy for modeling electrocapillary phenomena as well as the voltammetric response of electrodes in the presence of electrolytic environments.\cite{Karlberg2007, Bonnet2013} Yet, in spite of their remarkable success in describing hydrogen UPD on platinum, these models are difficult to apply when the adsorbates exhibit strong lateral interactions along the surface, as is the case for adsorbed transition metals. Reliable theoretical estimates of transition metal UPD adlayer stability thus remain challenging due to the complex nature of the interfacial structure, the critical influence of the applied voltage, as well as the need to account for configurational entropy to deliver accurate surface chemical potentials.

In this work, we present a quantum-continuum approach that addresses each of these challenges in turn, leading to an accurate description of metal adlayer stability. We treat solvent effects along the interface using the newly developed self-consistent continuum solvation (SCCS) model and describe an approach herein to model the effects of surface electrification.\cite{Andreussi2012} Using this method, we parameterize a cluster expansion of the electrified interface and perform grand canonical Monte Carlo (GCMC) calculations to obtain accurate adsorption isotherms that account for the configurational entropy along the surface.\cite{Tang2004} In order to introduce the method, we consider the UPD of silver on the gold (100) surface as it has been intensively studied over the years, and it has been shown to occur in a non-trivial three step process in both sulfuric and perchloric acid media.\cite{Garcia1994,Ikemiya1996,Garcia1998} 

\section{Computational Methods}
We model the deposition of silver by considering the equilibrium that exists between the silver ion and bulk silver 
\begin{equation}
\text{Ag}^+ + e^- \to \text{Ag},
\label{eq:bulk_silver}
\end{equation}
as well as the silver ion and the adlayer
\begin{equation}
\text{Ag}^+ + e^- + \ast \to \text{Ag}^\ast.
\end{equation}
The equilibrium between the surface and solution thus occurs when the chemical potential of the adsorbed silver is equal to the coupled chemical potential of the silver ion in solution and the electron in the electrode 
\begin{equation}
\mu_{\text{Ag}^\ast}(\theta, \Phi) =  \mu_{\text{Ag}^+} - e_0 \Phi.
\label{eq:equilibrium}
\end{equation}
Here, $\mu_{\text{Ag}^+}$ is the chemical potential of the silver ion in solution, which can be expressed in terms of the chemical potential of bulk silver and the formal reduction potential of the silver ion as indicated by Eq.~\ref{eq:bulk_silver}
\begin{equation}
\mu_{\text{Ag}^+} = \mu_{\text{Ag}}^\circ + e_0 \Phi_{\text{Ag}|\text{Ag}^+}.
\end{equation}
We calculate the formal potential of silver as $\Phi_{\text{Ag}|\text{Ag}^+} = \Phi_{\text{Ag}|\text{Ag}^+}^\circ + k_\text{B}T/e_0 \ln [\text{Ag}^+]$ V with respect to the standard hydrogen electrode (SHE) where $\Phi_{\text{Ag}|\text{Ag}^+}^\circ= 0.8$ V vs.~SHE is the standard reduction potential of silver and [Ag$^+$] is the bulk solution silver concentration.\cite{Haynes2016} We additionally define the surface chemical potential $\mu_{\text{Ag}^\ast}(\theta, \Phi)$ to have an explicit dependence on the surface coverage $\theta$ as well as the applied voltage $\Phi$. While the right hand side of Eq.~\ref{eq:equilibrium} can be computed directly at the level of DFT, the left hand side is considerably more challenging since the environment contributes non-negligibly to the energy of the adsorbed silver through solvation effects, surface electrification, as well as the lateral interactions amongst the neighboring atoms on the surface. 

Coverage and voltage effects on the stability of the silver adlayer are accounted for by performing quantum-continuum calculations of the metal-solution interface using planewave DFT as implemented in the \textsc{PWscf} code within \text{Quantum ESPRESSO} along with the SCCS model as implemented in the \textsc{Environ} module.\cite{Giannozzi2009, Andreussi2012, Dupont2013} The quantum electronic interactions are modeled with the Perdew-Burke-Ernzerhof exchange-correlation functional and the projector augmented wave method is used to represent the ionic cores. We found that kinetic energy and charge density cutoffs of 40 Ry and 480 Ry, respectively, yielded well-converged forces within 5 meV/\AA\ as well as total energies within 50 meV per cell. The Brillouin zone of each surface cell is sampled with a shifted $\frac {12}{n} \times \frac{12}{m} \times 1$ Monkhorst-Pack grid, so that the Brillouin zones of surface cells that consist of $(n\times m)$ primitive cells are consistently sampled. The electronic occupations are smoothed with 0.02 Ry of Marzari-Vanderbilt cold smearing.  Neutral surfaces are modeled within the slab-supercell approximation where the silver adlayers and the top and bottom two layers of a symmetric 7-layer gold (100) slab are allowed to relax. The slabs are centered in each cell and it was found that a vacuum height of 10 \AA\ was sufficient to converge the electrostatic potential at the cell boundaries using the recently implemented generalized electrostatic solvers in the module.\cite{Dabo2008, Andreussi2014} Solvent effects were modeled by replacing the vacuum region of the supercell with a polarizable continuum dielectric medium. The construction of the dielectric cavity was based on the parameterization of the SCCS model by Andreussi \emph{et al.} for neutral species.\cite{Andreussi2012} Non-electrostatic cavitational effects such as the solvent surface tension and pressure are additionally computed based on the quantum surface and quantum volume determined by the self-consistent shape of the cavity as described by Cococcioni \emph{et al}.\cite{Cococcioni2005} 

\section{Results and Discussion}
\subsection{Electrochemistry at charged interfaces}
\begin{figure}[t]
	\centering\includegraphics[width=0.95\columnwidth]{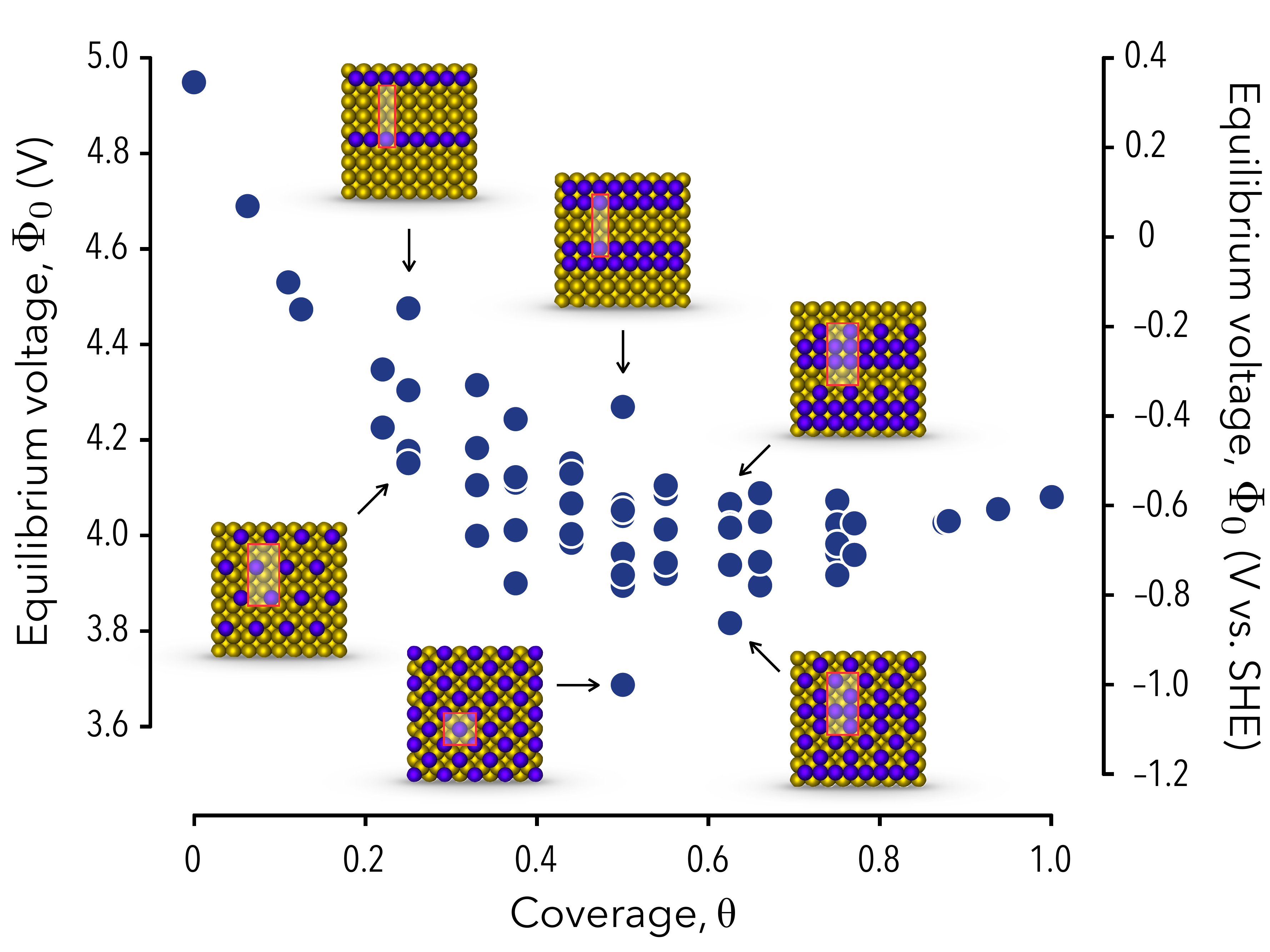}
	\caption{\small Equilibrium voltages extracted from the quantum-continuum calculations. Voltages on the absolute scale (left axis) are aligned to the SHE scale (right axis) by  recovering the experimental potential of zero charge of the bare surface. Noncompact (dispersed) configurations (lower insets) exhibit larger interfacial dipoles compared to compact (island forming) structures (upper insets).}
	\label{fig:equilibrium_voltages}
\end{figure}

Silver was found to adsorb preferentially in the hollow sites of the gold (100) surface. We sampled 59 different surface configurations with coverages spanning 0 - 100\% using surface cells ranging in size of $(1\times 1)$, $(2\times 2)$, $(2\times 4)$, $(3\times 3)$, and $(4\times 4)$ primitive surface cells (see supplemental section S1). The equilibrium voltage $\Phi_0$ of each neutral surface was computed by aligning the converged electrostatic potential to zero in the bulk of the solvent region, allowing us to extract voltages directly from the quantum-continuum calculations as the opposite of the Fermi level.\cite{Weitzner2017, Keilbart2017, Campbell2017} The equilibrium voltages were subsequently aligned to the SHE scale by ensuring that the potential of zero charge of the neutral bare gold (100) surface is aligned to the experimental value of 0.24 V vs.~SHE, as shown in Fig.~\ref{fig:equilibrium_voltages}.\cite{Kolb1986} We found that compact (island forming) configurations tended to have smaller interfacial dipoles than noncompact (dispersed) configurations. L\"{o}wdin population analysis revealed that charge transfers from the silver adlayer to the topmost gold layer and that the charge transfer occurs to a greater extent for the noncompact structures. This suggests that a stronger hybridization occurs between the orbitals of neighboring silver atoms on the surface than the hybridization that takes place between the valence orbitals of silver and gold. This charge transfer behavior additionally explains the initial increase in the interfacial dipole and the subsequent decrease beyond 50\% coverage.

\begin{figure}[h]
	\centering\includegraphics[width=0.95\columnwidth]{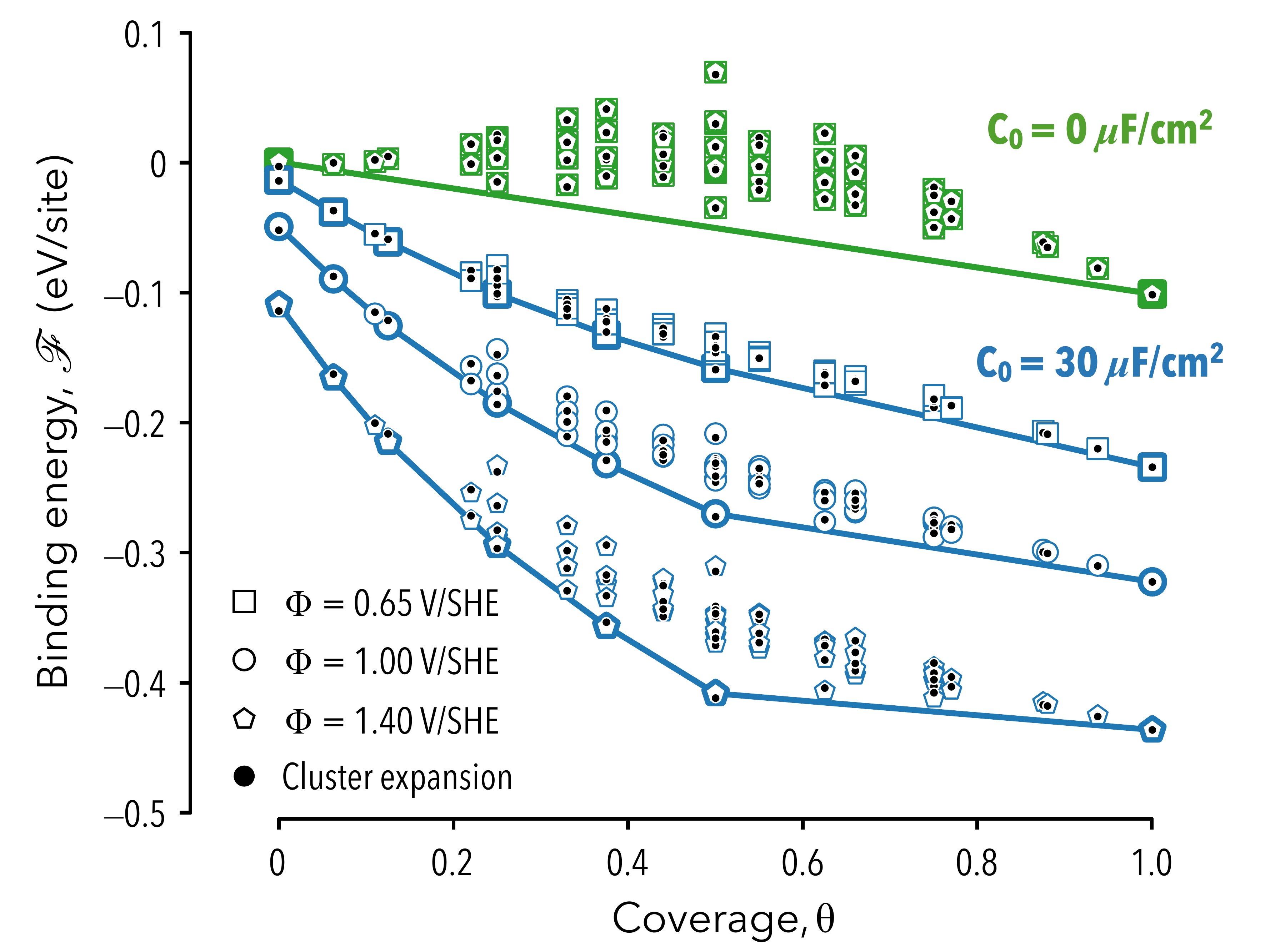}
	\caption{\small Effects of the differential capacitance on the voltage-dependent binding energies $\mathscr{F}(\bm{\sigma}, \Phi)$ for $C_0 = 0$ and $30$ $\mu$F/cm$^2$ (green/blue). The enhanced binding energy of intermediate coverages is driven by their large interfacial dipoles. Predicted energies from the cluster expansion are overlaid as solid circles. Ground state structures are identified with thicker markers. }
	\label{fig:formation_energies}
\end{figure}

 The occupancy of each site $i$ in the surface cells is represented by a spin variable $\sigma_i$, for which we adopt an Ising-like convention, where occupied sites are represented by a value of $+1$ and vacant sites by a value of $-1$. This enables us to describe a full configuration as a vector of spins $\bm{\sigma} = \{\sigma_i \}$. The binding energy of each neutral configuration was computed as
\begin{equation}
F_0(\bm{\sigma}) = \frac{1}{2} \Delta E(\bm{\sigma}) - N\mu_{\text{Ag}}^\circ,
\end{equation}
where $\Delta E(\bm{\sigma})$ is the difference in energies of a slab with configuration $\bm{\sigma}$ and the bare gold (100) surface, and $N$ is the number of occupied hollow sites on one side of the slab. Expanding the neutral binding energy with respect to the total charge $Q$ in the cell, we obtain the charge-dependent binding energy 
\begin{equation}
F(\bm{\sigma}, Q) = F_0  + \Phi_0 Q + \frac{1}{2}\frac{Q^2}{A C_0 },
\end{equation}
where $A$ is the area of one side of the slab and $C_0$ is the differential capacitance of the interface. The charge-dependent binding energy can be converted to a voltage-dependent representation by computing its Legendre transform with respect to the charge $\mathscr{F} = F - \Phi Q$, where $\Phi$ is the applied voltage. Here, the charge that develops on the surface at fixed voltage can be calculated as $Q = AC_0(\Phi - \Phi_0)$, directly capturing the effects of adsorption on the computed charge through the configuration-dependent potential $\Phi_0$. The differential capacitance of the interface is modeled by incorporating a Helmholtz plane into the solvent region of the supercell several angstroms from the surface. This capacitance can be computed directly with the quantum-continuum model yielding a range between 14 -- 21 $\mu$F/cm$^2$ (see supplemental section S2). We recognize however that the response of the physical double layer may exhibit a nonlinear dependence on the applied voltage and the concentration of the electrolyte. To take this dependence into account, we consider the differential capacitance to be an environmental parameter and perform a sensitivity analysis to assess its contribution to the overall stability of the silver monolayer, as demonstrated in Fig.~\ref{fig:formation_energies}.
%


In the case where the differential capacitance is set to $0\ \mu\text{F}/\text{cm}^2$, the binding energies are invariant with respect to the applied voltage. Furthermore, the only configurations that define the ground state of the system are the bare gold (100) surface and the full silver monolayer. This result is consistent with what would be found had these calculations been performed in vacuum and would similarly lead to the incorrect prediction that the monolayer formation occurs in one step or would appear as one peak in the voltammetry. However, accounting for a finite differential capacitance, we find that configurations with intermediate coverages become part of the ground state due to the large interfacial dipole associated with these configurations as shown in Fig.~\ref{fig:equilibrium_voltages}. This indicates that surface electrification controlled by the applied voltage directly influences the lateral interactions amongst the silver adatoms, as shown in Fig~\ref{fig:formation_energies}. We note that estimates of the surface chemical potential $\mu_{\text{Ag}^\ast}(\theta, \Phi)$ can be directly obtained from the binding energy-coverage curves as the slope of the common tangent lines connecting the configurations that lie on the ground state energy hulls. However, a key limitation of this approach is the missing configurational entropy that is needed to define accurate chemical potentials, as well as the fact that we have sampled only a small subset of the possible surface configurations leading to artificially discretized regions of stability.

\subsection{The voltage-dependent cluster expansion}
\begin{figure}[]
	\centering\includegraphics[width=0.95\columnwidth]{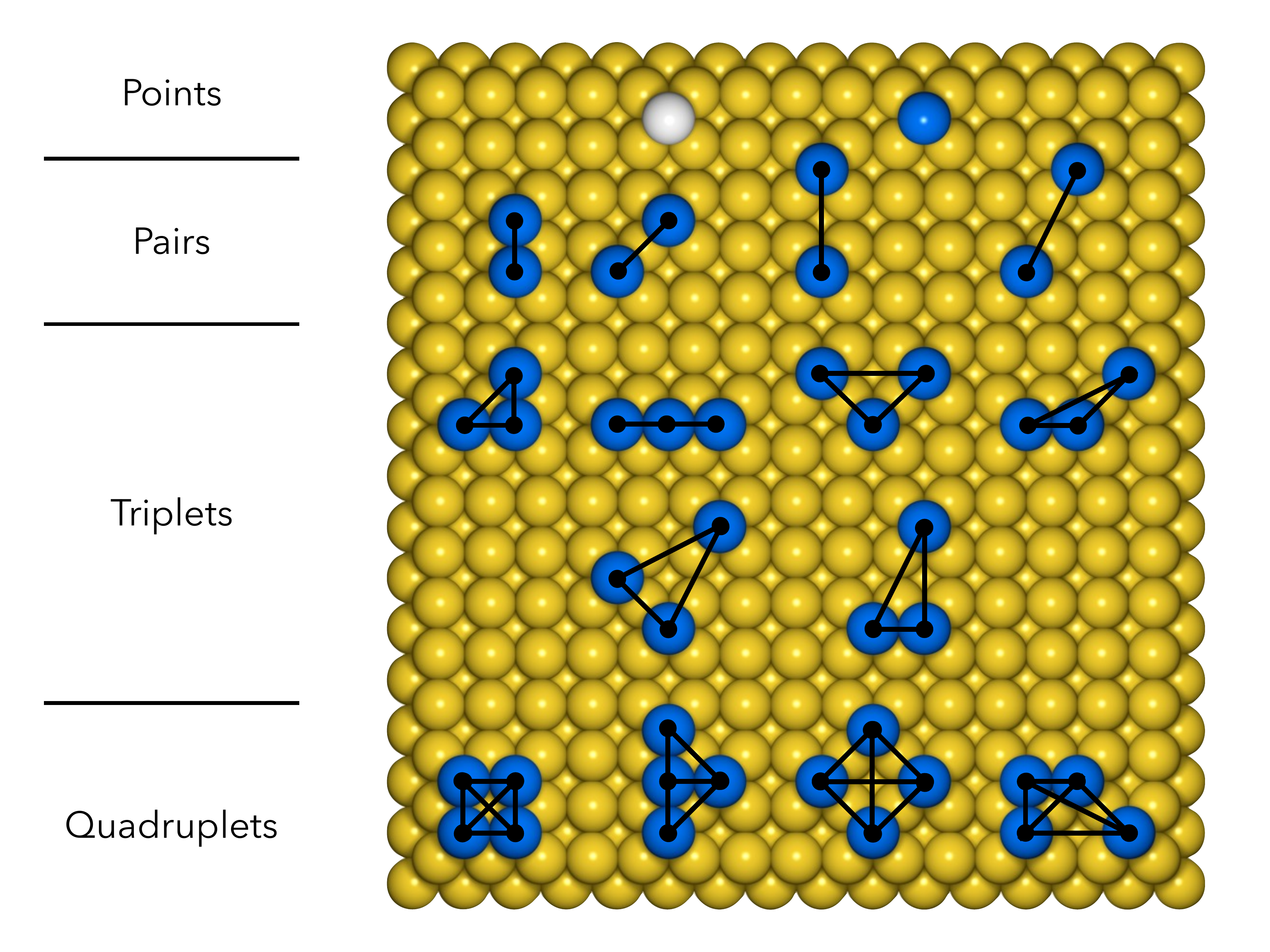}
	\caption{\small Clusters identified from the cluster selection process. Clusters with diameters that sample up to fourth nearest neighbors and cluster sizes up to quadruplets were included in the search. Sampled sites are shown in blue (except the empty point cluster shown in white).}
	\label{fig:clusters}
\end{figure}

In order to obtain an accurate chemical potential for silver on the gold (100) surface, we fit a cluster expansion to our dataset, enabling a rapid and accurate estimation of the voltage-dependent binding energy for considerably larger surface cells. The cluster expansion approach relies upon the construction of an infinite series expansion for which the expansion terms consist of polynomials of the spin variables $\sigma_i$.\cite{Sanchez1984} Each polynomial or \emph{cluster} of spins transforms under the symmetry operations of the underlying lattice, and as such, we refer to particular types or classes of clusters which we denote by $\alpha$ that is understood to belong to a set of symmetry related cluster functions. In practice, the expansion must be truncated and the cluster functions chosen in such a way so that only the most important clusters are retained. For a given set of clusters, the expansion is constructed by calculating the average of a cluster function of type $\alpha$ for a configuration $\bm{\sigma}$ as 
\begin{equation}
\bar{ \Pi }_\alpha(  \bm{\sigma}) = \frac{1}{m_\alpha M}\sum_{\beta \equiv \alpha } \prod_{ i \in \beta } \sigma_i,
\end{equation}
where $m_\alpha$ is a multiplicity factor equal to the number of clusters that are symmetrically equivalent to $\alpha$ related by the point group of the crystal, $M$ is the total number of sites in the lattice, and $i$ represents the site indices sampled by the cluster $\beta$.  The voltage-dependent binding energy per site of a configuration can then be computed as
\begin{equation}
\mathscr{F}(\bm{\sigma}, \Phi)/M = \sum_\alpha \bar{ \Pi }_\alpha m_\alpha J_\alpha,
\end{equation}
 where the $J_\alpha$ are the effective cluster interactions that ultimately determine the accuracy of the expansion. The effective cluster interactions are calculated via linear regression for a given expansion across the entire dataset. Candidate expansions are proposed following the formalism established in Ref.~\onlinecite{VandeWalle2002}, where a cluster may only be included if the expansion already contains its subclusters, and clusters that consist of $n$-sites of a certain diameter may only be included if all $n$-site clusters of a smaller diameter are already present. The set of candidate expansions considered in this work can be found in section S3 of the accompanying supplemental document. Cluster selection is then carried out by performing leave-one-out cross validation analysis (LOOCV) for all possible clusters that sample up to four sites (quadruplets) and have a maximum diameter of up to fourth nearest neighbors. LOOCV provides a score $\Delta$ ranking the ability of a given expansion to accurately predict configurational energies. The score is calculated as 
\begin{equation}
\Delta =  \left( \frac{1}{k} \sum_{i=1}^k ( \mathscr{F}_i - \hat{ \mathscr{F} }_i )^2 \right)^{\frac{1}{2}},
\end{equation}
where $\mathscr{F}_i$ is the energy of configuration $i$, and  $\hat{\mathscr{F}}_i$ is the predicted energy of configuration $i$ from a linear fit to the other $k-1$ configurations in the dataset. In this work, we have identified a basis set consisting of sixteen clusters depicted in Fig.~\ref{fig:clusters} that provides an accurate description of the ground state with a LOOCV score between 1.8 and 14.6 meV/site in the considered voltage range for differential capacitance values between 0 and 100 $\mu$F/cm$^2$.
The voltage-dependent binding energies predicted by this model for differential capacitances of 0 and 30 $\mu$F/cm$^2$ are shown in Fig.~\ref{fig:formation_energies} at voltages of $0.65$, $1.00$ and $1.40$ V/SHE, demonstrating the evolution of the ground state between the lower and upper bounds of voltages considered in our analysis. We find that across the entire considered voltage range, small compact clusters with diameters less than two nearest neighbors contribute the most significantly to the binding energy of a given adlayer suggesting the importance of short range correlation effects to the adlayer stability (see supplemental section S3 for more details).

\begin{figure}[h]
	\centering\includegraphics[width=0.95\columnwidth]{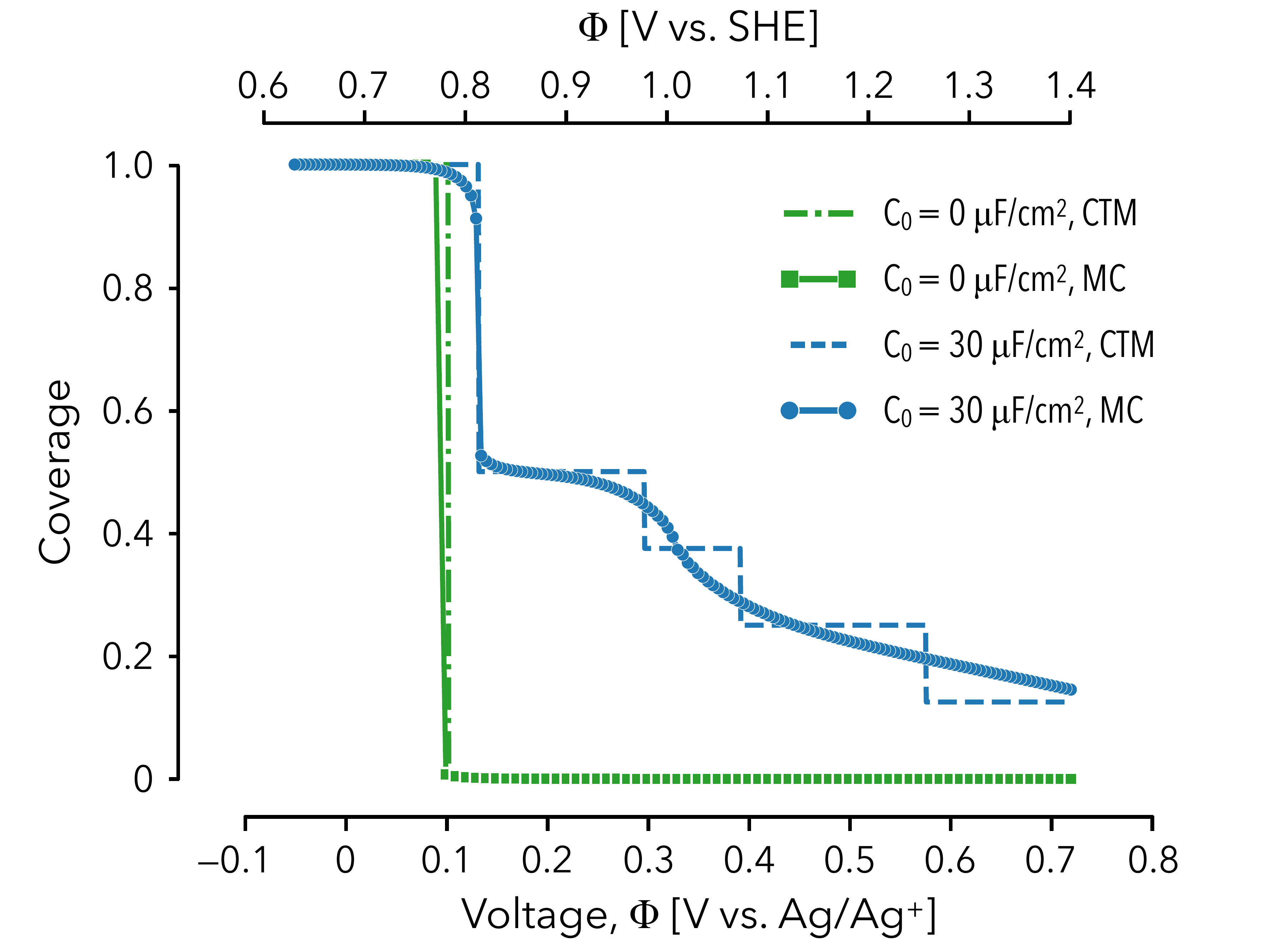}
	\caption{\small Theoretical adsorption isotherms obtained for a bulk solution silver concentration of [Ag$^+]=10^{-2}$ M. Isotherms were obtained using both the Monte Carlo (MC) and common tangent method (CTM) for differential capcitance values of 0 and 30 $\mu$F/cm$^2$. Coverages were averaged over 20,000 Monte Carlo steps after 5,000 Monte Carlo steps of warm up with standard deviations lower than $5\times 10^{-2}$.}
	\label{fig:isotherms}
\end{figure}

Using the cluster expansion as a model Hamiltonian, we perform GCMC calculations of the interface using the Metropolis-Hastings algorithm. The grand potential of the system can be expressed as $ \phi(\mu_{\text{Ag}^+ + e^-}, \Phi) = \mathscr{F}(\bm{\sigma}, \Phi) - N\mu_{\text{Ag}^+ + e^-}$, where $\mu_{\text{Ag}^+ + e^-}$ is the coupled chemical potential of the silver ion and electron, as in the right hand side of Eq.~\ref{eq:equilibrium}. We determine the equilibrium surface coverage over a range of voltages for a cell that consists of $20\times20$ primitive surface cells. Each trajectory is initialized with a random coverage of 50\% and allowed to warm up for 5,000 Monte Carlo steps prior to collecting data for averaging over the course of 20,000 Monte Carlo steps. Applying this methodology, we have calculated adsorption isotherms for the UPD of silver on the gold (100) surface as shown in Fig.~\ref{fig:isotherms}, which we compare to isotherms obtained by applying the common tangent method to the binding energies shown in Fig.~\ref{fig:formation_energies}.
Here we observe that when the differential capacitance is taken to be 0 $\mu$F/cm$^2$, the silver monolayer appears to form in one step, as expected from its binding energy curve. Furthermore, entropic and voltage effects influence the shape of the isotherms negligibly. In contrast, for a differential capacitance of 30 $\mu$F/cm$^2$, the isotherms exhibit multiple transitions in the surface coverage as a direct result of accounting for the variation of the interfacial dipole. Moreover, the effects of configurational entropy and the enhanced sampling afforded by the cluster expansion and the GCMC provide an accurate description of the voltage-dependent interfacial equilibria as compared to the discretized isotherm directly obtained from the binding energies of the underlying dataset. We note that surface electrification alone can elicit multiple transitions in the surface coverage, confirming the importance of the excess surface charge in describing the deposition process as suggested by Ikemiya, Yamada, and Hara.\cite{Ikemiya1996} The results obtained herein may additionally be improved upon by introducing co-adsorption effects into the model, as it has been indicated that an adjacent layer of (bi)sulfate or perchlorate may be present at the interface throughout the deposition process.\cite{Garcia1998} It is well known that the presence of co-adsorbates can strongly alter the composition and structure of alloy surfaces, and may play an important role in metal monolayer formation.\cite{Han2005,Gimenez2010,Velez2012} The introduction of co-adsorption effects combined with the voltage-dependent cluster expansion proposed herein is expected to provide a powerful computational treatment of underpotential deposition and related heterogeneous processes.

\section{Conclusion}
In summary, we have developed an embedded quantum-continuum model of electrodeposition phenomena that accounts for the configuration-dependence of the interfacial dipole. The methodology was presented by considering the underpotential deposition of silver onto the gold (100) surface due to the complexity associated with the silver monolayer formation process. We demonstrated how voltage-dependent binding energies can be computed for an array of surface configurations to  parameterize a cluster expansion of the interface. Grand canonical Monte Carlo calculations of the interface highlighted the critical need to account for the interfacial dipole as well as entropic effects when modeling the stability of deposited metals. The method presented in this work is widely applicable to the design of shaped transition metal/alloy nanoparticles, and may be useful in the design of nanostructured catalysts and nanoparticle-based optical sensors.

The authors acknowledge primary support from the National Science Foundation under Grant DMR-1654625, and partial support from the Center for Dielectrics and Piezoelectrics at Penn State University. The authors thank the Penn State Institute for CyberScience for providing high-performance computing resources and technical support throughout this work. This work used the Extreme Science and Engineering Discovery Environment (XSEDE), which is supported by National Science Foundation grant number ACI-1548562.


%

\end{document}